\documentclass{article}
\usepackage{arxiv}
\usepackage[utf8]{inputenc}
\usepackage{subcaption}
\usepackage[export]{adjustbox}[2011/08/13]
\usepackage{natbib}
	\setcitestyle{authoryear,round,aysep={}}
\usepackage{float}
\usepackage[linesnumbered, ruled, vlined]{algorithm2e}

\usepackage{amssymb}
\usepackage{pifont}
\newcommand{\cmark}{\ding{51}}%
\newcommand{\xmark}{\ding{55}}%

\SetKwComment{Comment}{$\triangleright$\ }{}

\SetCommentSty{mycommfont}
\usepackage{enumitem}

\usepackage[normalem]{ulem}

\usepackage{appendix}
\usepackage{booktabs}
\usepackage{multirow}
\usepackage{array}
\newcolumntype{L}[1]{>{\raggedright\let\newline\\\arraybackslash\hspace{0pt}}m{#1}}
\newcolumntype{C}[1]{>{\centering\let\newline\\\arraybackslash\hspace{0pt}}m{#1}}
\newcolumntype{R}[1]{>{\raggedleft\let\newline\\\arraybackslash\hspace{0pt}}m{#1}}
\usepackage[table]{xcolor} 
\usepackage{graphicx}

\usepackage{mathtools}

\usepackage{amsmath}
\usepackage{cleveref}
\usepackage[most]{tcolorbox}

\newenvironment{concepts}{
\begin{tcolorbox}[title=Relevant Concepts \thesubsection]
}{\end{tcolorbox}}

\newtcbtheorem[auto counter,number within=section]{concepts2}
  {Relevant Concepts}{}{theorem}

\title{Modelling Human Routines: Conceptualising Social Practice Theory for Agent-Based Simulation}
\author{Rijk Mercuur, Virginia Dignum, Catholijn M. Jonker}
\date{}

\begin{document}
\maketitle
\begin{abstract}
Our routines play an important role in a wide range of social challenges such as climate change, disease outbreaks and coordinating staff and patients in a hospital. To use agent-based simulations (ABS) to understand the role of routines in social challenges we need an agent framework that integrates routines. This paper provides the domain-independent Social Practice Agent (SoPrA) framework that satisfies requirements from the literature to simulate our routines. By choosing the appropriate concepts from the literature on agent theory, social psychology and social practice theory we ensure SoPrA correctly depicts current evidence on routines. By creating a consistent, modular and parsimonious framework suitable for multiple domains we enhance the usability of SoPrA. SoPrA provides ABS researchers with a conceptual, formal and computational framework to simulate routines and gain new insights into social systems. 
\end{abstract}

\section{Introduction}
Our routines play an important role in a wide range of social challenges such as climate change, disease outbreaks and coordinating staff and patients in a hospital.  Studying these systems via agent-based simulations enables researchers to gain insight into complex aspects of these challenges such as human interaction, individual adaptability, heterogeneity, feedback loops and emergence. To use agent-based simulations to understand the role of our routines in social challenges we need an agent framework that integrates our routines \citep{Mercuur2020}. \citet{Mercuur2020} recognizes this and argues for grounding such an agent framework in social practice theory (SPT). Social practice theory is a socio-cognitive theory that emphasizes how humans use their routines (practices) to come to a common view of the world (a social view). Using a socio-cognitive theory, such as SPT, enables ABS researcher to reuse evidence from the social sciences and relate the agent framework to other social theories. In short, translating SPT to an agent framework supports researchers in simulation studies on our routines that are grounded in evidence and embedded in the social sciences.

\begin{figure}[ht!]
    \centering
    \includegraphics[width=0.5\textwidth]{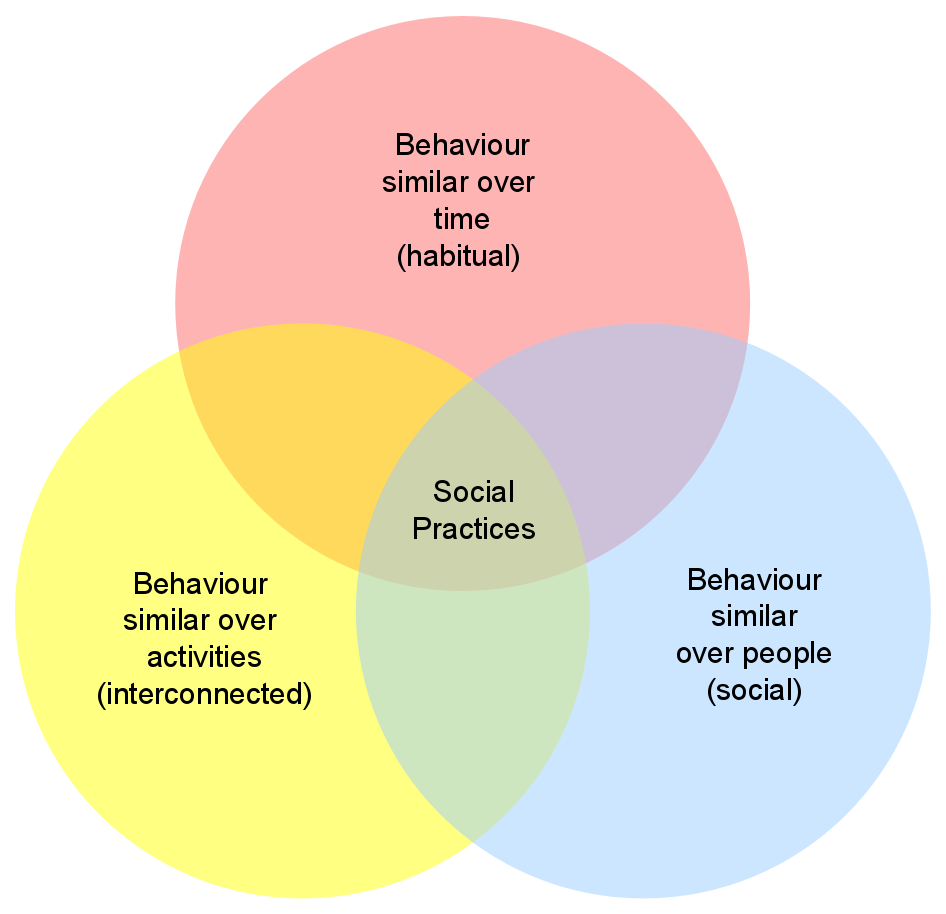}
    \caption{Social practices describe behaviour that is similar over time, activities and people (adapted from \citep{Mercuur2020}).}
    \label{fig:venn}
\end{figure}

To further an agent framework that integrates SPT, \citet{Mercuur2020} identifies relevant aspects of SPT and establishes a set of requirements for integrating SPT in agent-based models (ABM). They identify that SPT gives insight into the habitual, social and interconnected nature of our routines (see Figure \ref{fig:venn}). For example, the SP of commuting consists of a series of interconnected actions that are chained habitually: getting in the car, driving the kids to school and getting to work. These practices --  chains of actions -- are used in social situation as a mental model to reason about others actions (e.g., to coordinate the carpool). The authors distill requirements for integrating these aspects from the literature on agent theory, social psychology and SPT.  For example, a requirement related to this example states that agents should use social practices as a collective view on the world to coordinate their actions (see Appendix \ref{app:req} for all requirements). The authors evaluate current agent frameworks and conclude that -- although these frameworks are useful for their own purposes -- they do not satisfy the requirements for modelling our routines. In short, a new framework is needed that integrates SPT in agent models and satisfies the requirements set out by \citet{Mercuur2020}.

This paper extends current work that integrates SPT in agent models. \citet{Narasimhan2017} used SPT in an ABS to study a specific domain: energy systems. In contrast, our paper emphasizes reusability by providing a \emph{systematic} translation from theory to framework to enable a \emph{domain-independent} framework. \citet{Holtz2014} studies SPT via ABS by studying SPs as if they were agents themselves. In contrast, our work integrates SPT and agent theory enabling insights in the interaction of humans and SPs. Most notably for our purpose is the high-level conceptual model of SPs by \citet{DignumSP} and the related formalization in the pre-print \citet{DignumFormalSP}. \citet{DignumFormalSP} presents a formalization that integrates SPT and agent theory with a different scope, audience and resulting view on SPs than ours. \citet{DignumFormalSP} nevertheless serves as an essential inspiration and starting point for our framework. As such, we consider the concepts \citet{DignumSP} uses in our own framework and relate our framework to theirs in the discussion. 

This paper provides the domain-independent social practice agent (SoPrA) framework that satisfies the requirements set out by \citet{Mercuur2020}. We approach this by using concepts from the literature on agent theory, social psychology and SPT described in \citet{Mercuur2020}. For each modelling choice, we present an overview of the relevant concepts using delineated boxes. We describe how based on these concepts we provide a new conceptualisation (i.e., new labels and a new organization) that fulfils our requirements while aiming for reusability (i.e., compactness and modularity). The concepts from the literature are marked as either (1) directly represented as a concept in the UML (\cmark), (2) (partly) subsumed by the concepts in our UML ($\simeq$) or (3) left out the UML because they are not necessary to fulfill our requirements (\xmark). Our UML provides constrains in how the chosen classes relate by depicting associations, the multiplicity of the associations and the types of the attributes. For example, our UML shows locations are not also resources as objects cannot be a member of two distinct classes. In short, by building upon concepts of the literature we provide a clear and correct framework in UML.

SoPrA is implemented in UML, OWL and Java.  The main body of this paper describes our framework in UML. UML is a standardised, popular language that makes the framework easier read, understand, code, extend, maintain and adapt \citep{Bersini2012}. We show the UML adheres to the complete set of requirements and correctly implements them. The implementation of the requirements is by design, i.e. each subsection argues for a specific framework that embodies the requirement. Our online repository provides a formal framework in the OWL-language, which shows our framework is consistent, and a computational framework in Repast Java, which enables simulating social phenomena with the framework (see Appendix \ref{app:impl}). These frameworks provide ABS researchers with reusable, consistent and computational framework to simulate our routines and gain new insights into social systems.

The remainder of this paper is structured as follows. Section \ref{sec:conceptualising:highlevelmodel} presents the high-level choices made to connect SPT and agent theory. Section \ref{sec:conceptualising:habituality}-\ref{sec:conceptualising:interconnectivity} provides the primary concepts and relations to researchers to model respectively habituality, sociality and interconnectivity. Section \ref{sec:conceptualising:discussion} discusses the resulting framework by exemplifying the new insights enabled via simulation and the criteria on which we made model choices. Furthermore, Section \ref{sec:conceptualising:discussion} concludes that our paper provides ABM researchers with a computer model to simulate our routines and gain new insights into social systems. Our paper ends with three appendices: Appendix \ref{app:req} describes the requirements set out by \citet{Mercuur2020}, Appendix \ref{app:ext} provides possible extensions of SoPrA. Appendix \ref{app:impl} describes the computational and formal versions of SoPrA.

\newpage
\section{High-Level Modelling Choices}
\label{sec:conceptualising:highlevelmodel}
\begin{figure}[ht]
    \centering
    \includegraphics[width=0.8\textwidth]{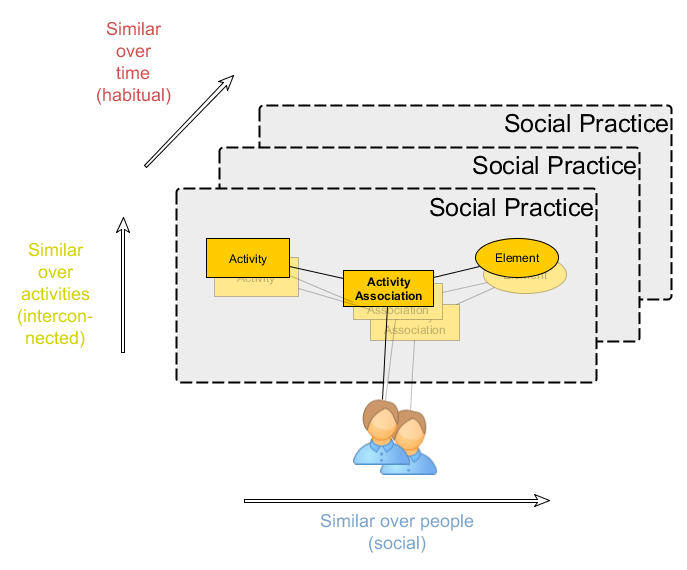}
    \caption{A high-level overview of SoPrA depicting the main concept -- activities, elements and agents -- and the ternary connection between these concepts that can be similar over time, people and activities.}
    \label{fig:sopra-simple-similarities}
\end{figure}

SoPrA connects the collective perspective on behaviour of SPT and the individual perspective of agent theory. To capture the collective perspective of SPT, SoPrA should capture the similarity of behaviour over time, people and activities. To capture the individual perspective of agent theory, SoPrA should provide researchers with the primary concepts and relations to model agents that make (1) habitual decisions, updates and reason about habituality, (2) socially intelligent decisions, updates and reason about collective concepts and (3) interconnected decisions, updates and reason about the interconnectedness of activities (H2, S2, I2). Figure \ref{fig:sopra-simple-similarities} provides a high-level overview of the concepts and relations we use to connect these two perspective: activities, elements, agents and activity-associations. The remainder of this section explains how these concepts connect SPT and agent theory. 

Activities refer to the bodily movement itself without all the mental connotations. In SPT there are two views on the relation between activities and SPs. First, according to \citet{Reckwitz2002} an SP represents a pattern which can be filled out by a multitude of single and often unique activities reproducing the SP. An SP thus consists of several activities. For example, the SP of commuting can consist of taking your kids to school and then going to work.  Second, for \citet{Shove2012} activities are not a part, but instead another view on the SP. They view the SP as either a collection of elements (named practice-as-entity) or as something that is performed (named practice-as-performance). For example, they view commuting as either a collection of elements, such as the meaning of transport and the material car, or as a series of activities, such as getting in your car and going to work. We follow \citet{Reckwitz2002} in that activities are elements that are part of an SP (and not different views on an SP) because this matches the agent perspective where activities are also part of an agent model (and not different views on an agent model). In agent theory, activities (or actions, see \ref{sec:conceptualising:interconnectivity}) are the result of a decision-making process. For example, an agent chooses to move a block, commute by car or sell a stock. In short, activities are part of both SPs and agent models and therefore enable us to connect both perspectives.

Elements refer to epistemological entities such as places, human values or resources. In agent theory, these elements are part of the mental model of the agent in the form of typical agent constructs such as goals, desires or resources. In SPT, the aforementioned practice-as-entity perspective entails that SPs are collections of interconnected elements. \citet{Taylor1973} first describes the SP as an entity separate from an agent with multiple elements: ``meanings and norms implicit in [...] practices are not just in the minds of the actors but are out there in the practices themselves". SPs thus comprise other elements than activities (e.g., meanings and norms). \citep{Reckwitz2002} describes the connection between these elements and the SP as that the existence of SPs depends on the existence and specific organization of these elements and cannot be reduced to any single one of them. For example, the SP of skateboarding depends on a specific group of elements that interconnect: materials like the skateboard or street spaces, the competences to ride the board, rules and norms like what defines a trick, and the meaning different groups attribute to the practice like recreation or transport \citep{Shove2012}. In short, elements are part of both SPT and agent theory. This paper describes elements described in SPT and agent theory and chooses which elements suffice to conceptualize an agent model that integrates SPT (and reflects the requirements).

Agents refer to the individual decision-makers that represent humans. Whereas in agent theory the agent is a central component in SPT the agent is subsidiary. In SPT, the social practice `recruits' agents as `hosts' and uses them as a vehicle to spread \citep{Shove2015}. In fact, one common motivation to use SPT is to be able to abstract away from the actor \citep{Shove2015, Reckwitz2002, Schatzki1996}. This paper acknowledges the strength of SPT in expressing this collective perspective, but to serve ABM aims to connect this perspective with the individual perspective of agent theory.

Activity associations refer to the mental connection between an activity an element. In agent theory, agents make mental connections between activities and other mental constructs. For example, BDI-agents connect their desire $D$ to activity $A$ by having the belief that $A$ satisfies $D$. In SPT, activity associations are called connections; they connect an activity and another element of an SP \citep{Bourdieau1977,Reckwitz2002,Shove2012}. In SoPrA, activity associations are the central component that connects SPT and agent theory. Figure \ref{fig:sopra-simple-similarities} shows how activity associations connect activities, elements and agents.

SoPrA associates elements and agents (e.g., meanings, materials) with specific activities and not --- as is common in SPT literature \citep{Bourdieau1977,Reckwitz2002,Shove2012} --- with the SP as a whole. For example, an agent does not associate the SP of commuting with environmentalism. Instead, an agent associates an activity instantiating commuting --- taking the train to work --- with environmentalism, while associating another instance--- taking the car to work --- with the meaning of efficiency. We give four reasons for this modelling choice.
First, this enables modelers to capture the beliefs an agent has about an SP in a nuanced way (e.g., make a difference between forms of commuting that promote different values).
Second, it enables modellers to use these nuances to guide the decision-making of the agent (e.g., choose car-driving over the train because it's more efficient). 
Third, it enables modellers to feed SoPrA with a selection of facts about actions and let the agent make inferences about composite or sequential activities (e.g., commuting is boring, because all of its implementations are boring). 
Fourth, it corresponds to how empirical work in neurology conceptualizes human decision-making \citep{Metzinger2003}. In \citet{Metzinger2003}, associations between activities and goals (i.e., one of the SP elements) are used both to reason about our own decisions as well as a way to reason about others (i.e., social intelligence). 
In short, associating elements with specific activities -- instead of with the SP as a whole -- enables modellers to capture the nuances within a SP, use these nuances to guide decision-making, use a circumscribed knowledge base and corresponds to evidence within neurology.

Figure \ref{fig:sopra-simple-similarities} presents how our requirements relate to the choices made in conceptualising an agent model that integrates SPT: a focus on activities as part of a SP, conceptualising SPs as a collection of elements and emphasizing the associations between activities and elements. These choices enable us to express more clearly what it means for behaviour to be similar over time, people and activities (requirement 1, 4, 6). The similarity of behaviour lies in activity associations. Thus habituality captures that people associate the same SP elements with an activity over time; sociality captures that different people associate the same elements with an activity; interconnectivity captures that people associate the same elements with different activities. SPs thus capture that activity associations are similar over time, people and different activities. 

\begin{figure}[ht]
    \centering
    \includegraphics[width=\textwidth]{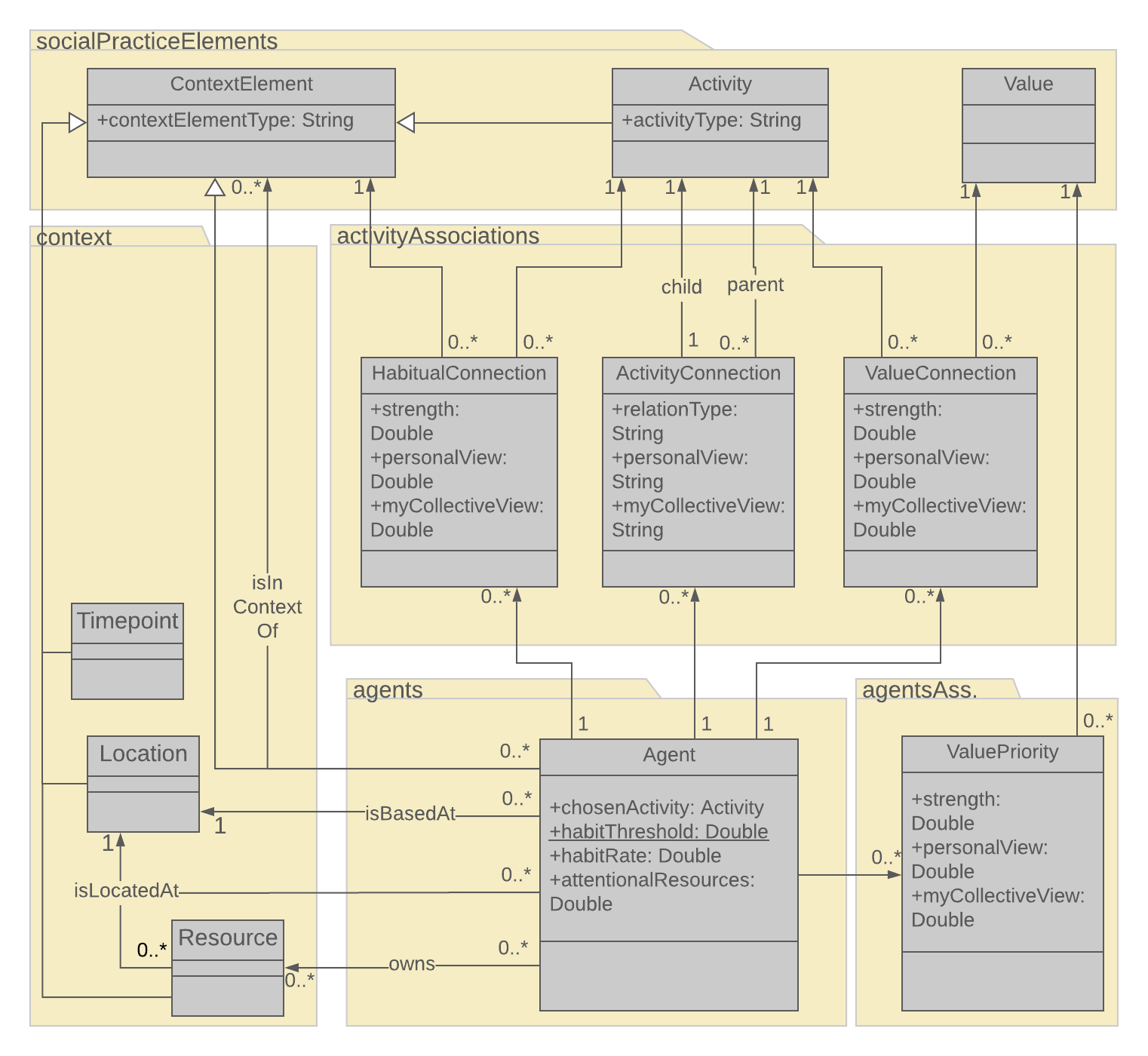}
    \caption{SoPrA in UML that integrates social practie theory and agent theory emphasizing habituality, sociality and interconnectivity.}
    \label{fig:uml}
\end{figure}

This leaves open the question of how we make these concepts and relations precise to enable agents to reason about habits, sociality and interconnectivity (H2, S2, I2). Figure \ref{fig:uml} presents an overview of the full UML model with the specific concepts and relations that suffice to implement the requirements. The remainder of this paper refers to this figure and explains why these concepts and relations model habitual, social and interconnected behaviour. 

\newpage
\section{Conceptualising Habituality}
\label{sec:conceptualising:habituality}
SoPrA provides the primary concepts and relations to researchers to model habitual decisions, updates and reasoning (H.2). The more concrete requirements on habits (H3-H7) specify in detail what is needed to model habits. For example, agents need to be able to differentiate between habits and intentions (H3). Figure \ref{fig:uml} shows that agents are able to differentiate between a habitual association (i.e., \texttt{HabitualConnection}) and an intentional association (i.e., \texttt{ValueConnection}). This leaves open how we conceptualize what triggers habits (H4), what concept captures habitual triggers (H5), how we enable agents to differ in their habits (H6) and why intentions are directed at an abstract value (H7). The following subsections treat these aspects in more detail.

\subsection{Context-Elements}
Our conceptualisation of context-elements captures that context-elements can comprise resources, activities, location, timepoints and other people (H.5). Aside, the conceptualisation  of context-elements should use comprehensive set of collective concepts needed for social reasoning (S3).\footnote{Physical objects namely play a central role in our social world in that they provide a collective common ground \citep{Searle1995}.} Box \ref{box:ce} gives an overview of the concepts relevant to conceptualize the \texttt{context} package that fulfills requirement H.5 and S.3. The remainder of this subsection explains SoPrA.


\begin{concepts}
\begin{description}
\item[Material$^\simeq$] Materials refer to the physical aspects of a SP \citep{Schatzki1996,Latour1996,Reckwitz2002,Shove2012} . For \citet{Shove2012} this comprises the resources people relate to the SP. For example, the SP of commuting relates to a car, bike, or public transportation.
\item[Context$^\simeq$]] Context is any information that can be used to characterise the situation of an entity. An entity is a person, place, or object that is considered relevant to the interaction between the user and the application, including the user and the applications themselves \citep{Dey2001}.
\item[Context Cue$^\text{\cmark}$] A contextual cue is any event noticed by the organism, with the exclusion of the target stimuli that form the learning experience (Learning and Memory: A Comprehensive Reference, 2008)
\item[Location$^\text{\cmark}$] A location is a fixed geographical point.
\item[Resource$^\text{\cmark}$] A resource describes a physical entity that has some utility.
\item[Time-Point$^\text{\cmark}$] A fixed temporal point. For example, morning, afternoon, 12pm or Wednesday.
\end{description}
\label{box:ce}
\end{concepts}

To satisfy requirement H.5 we model a \texttt{ContextElement} class that generalizes the \texttt{Activity} class, \texttt{Agent} class, \texttt{Timepoint} class, the \texttt{Location} class and the \texttt{Resource} class. The UML clarifies the relation between these classes (e.g., a resource cannot also be a location as UML forces these objects in seperate classes). The \texttt{ContextElement} class has an attribute \texttt{contextElementType} that can be assigned to the string \texttt{"{}Activity"{}}, \texttt{"{}Agent"{}}, \texttt{"{}Location"{}}, \texttt{"{}Resource"{}} or \texttt{"{}Timepoint"{}} such that an agent can easily differentiate between the types of context-elements. We chose the wording context-element instead of context cue, because these objects have more function in the model than `cueing' a habit. The concept of `material'  is entailed by the concept of `context-element' (in particular, by the class \texttt{Resource} it generalizes) and therefore does not require a separate class. Locations, resources, timepoints, agents and activities each trigger habitual behaviour, but have additional, but different, functions in the model and therefore are modellled as seperate classes. This conceptualisation of context-elements thus captures all necessary elements that trigger habits and still enables these elements to have different functions in the model.

SoPrA models context-elements as tokens -- not types. The type-token distinction refers to the difference between a general sort of thing and its concrete instances \citep{sep-types-tokens}. For example, the token \texttt{Car1} is a possible context-element, but the type \texttt{Car} is not. Usually a Class in UML refers to a type and an instance of the class to a token. However, because we aim for a domain-independent framework we abstract over multiple types of context-elements (cars, restaurants, homes) for which different tokens trigger different habits. For example, a modeller should separate between \texttt{BobsCar} and \texttt{AlicesCar}, the former triggering a habit for Bob while the latter does not. Note that someone else' car, to some extent, also triggers your car driving habit. Modellers capture this by modelling a context-hierarchy that relates a more abstract context-element to a more particular one. For example, to model a class `Car' that has multiple instances `Bobscar' and `AlicesCar'. SoPrA thus models context-elements as tokens to provide the fine-grained static groundwork on which, depending on the target system, researchers can build more complex context hierarchies.

\subsection{Habitual Connections}
SoPrA supports habitual connections between an action and a context-element where the strength of that relationship is a continuous parameter (H.4). Box \ref{box:triggers} gives an overview of the concepts relevant to conceptualize the \texttt{HabitualConnection} class that fulfills requirement H.4. The remainder of this subsection explains SoPrA.

\begin{concepts}
\begin{description}
\item[Connection$^\text{\cmark}$] The relation between the elements of a social practice. For example, the connection between a commuting activity, material car and the value of environmentalism.
\item[Habitual Trigger$^\simeq$] The cue in the context that triggers the habit \citep{Neal2012}.
\item[Habit Strength$^\simeq$] The extent to which an action is experienced as habitual. There are several methods for measuring habit strength: the response frequency measure (RFM) \citep{Verplanken1994}, behavioural frequency \citep{Wood2002} and the self-reported habit index (SRHI) \citep{Verplanken2003}.
\end{description}
\label{box:triggers}
\end{concepts}

The \texttt{HabitualConnection} class associates an \texttt{Activity}, \texttt{ContextElement} and an \texttt{Agent}. The class captures the mental connection between a context-element, an activity and an agent. In contrast to a habitual trigger it denotes the relation and not the context-element itself. The \texttt{strength} attribute of \texttt{HabitualConnection} class represents to what extent the context element is connected with the activity. The stronger these associations the more chance these context-elements will trigger a habit. In social psychology, habit strength is measured on different levels of specificity: per activity (across different context) (RFM\citep{Verplanken1994} and the SRHI \citep{Verplanken2003}) or per activity and per context \citep{Wood2002}. In SPT, a connection relates a general activity to a specific element. SoPrA provides the fine-grained groundwork underlying each of these conceptualisations: we denote the habitual strength of a mental connection between a specific activity (e.g., commuting by car instead of the more general commuting) and a specific context-element (e.g., my front door instead of the more general my home) . 


\subsection{Heterogeneity in Agents' Habitual Tendencies}
SoPrA supports modelling agents that differ in (1) the strength of a habitual connection, (2) the maximum strength of a habitual connection, (3) the time to reach this maximum and (4) the amount of attention they attribute to a decision (H.6). Box \ref{box:attention} gives an overview of the concepts relevant to conceptualize the \texttt{Agent} class that fulfills requirement H.6. The remainder of this subsection explains SoPrA.

\begin{concepts}
\begin{description}
\item[Attention$^\simeq$] The allocation of cognitive resources among ongoing processes \citep{Anderson2005}.
\item[Attentional Resources$^\text{\cmark}$] The limited amount of cognitive resources available at one moment of time \citep{Anderson2005}.
 \citep{Anderson2005}. 
\end{description}
\label{box:attention}
\end{concepts}
The \texttt{Agent} class contains an \texttt{attentionalResources} attribute that captures the current amount of attentional resources available to allocate to a decision. The amount of attention attributed to a decision is the product of a number of cognitive processes using the same attentional resources \citep{Anderson2005}. In line with this thought, we chose to model the \texttt{attentionalResources} attribute as an integer.

The \texttt{Agent} class contains a \texttt{habitRate} attribute that ensures heterogeneity in how fast agents acquire habits. This attribute captures how much the \texttt{strength} of the \texttt{HabitualConnection} class increases when an agent experiences an action in relation to a \texttt{ContextElement}. As shown in \citep{Mercuur2017}, a difference in \texttt{habitRate} suffices to acquire heterogeneity in both the learning rate as well as the maximum habit strength agents report. 

The \texttt{Agent} class contains a \texttt{habitThreshold} attribute that captures a cut-off point: if the agent experiences an habitual pressure above the habit threshold it will make a habitual decision instead of an intentional decision. The habit threshold is the same for every agent as \texttt{habitRate} and \texttt{attentionalResources} suffice to acquire heterogeneity in the inclination of agents to fall into a habit.

\subsection{A Teleological Construct}
\label{sec:conceptualising:habits:teleo}
SoPrA provides a concept that captures the abstract aim agents direct their intentions at (H.7). Box \ref{box:teleo} gives an overview of the concepts relevant to conceptualize the \texttt{Value}, \texttt{ValuePriority}, \texttt{ValueConnection} class that fulfills requirement H.7. The remainder of this subsection explains SoPrA.

\begin{concepts}
\begin{description}
\item[Meaning$^\simeq$] Symbolic meanings, ideas and aspirations associated with a SP such as, attributing the meaning of `health' to the SP of riding the bike \citep{Shove2012}.
\item[Goal$ ^\text{\xmark}$] A goal is an idea of the future or desired result that an agent commits to achieve \citep{locke1990theory}. 
\item[Desire$ ^\text{\xmark}$] Represent the motivational state of the system \citep{BDI}.
\item[Value$^\text{\cmark}$] Represent what one finds important in life (Poel \& Royakkers 2011). For example, privacy, wealth or fairness.
\item[Social Motives$ ^\text{\xmark}$] Basic natural incentives that give rise to “energizing” subsequent action . In paricular, (1) achievement, (2) power, (3) affiliation and (4) avoidance \citep{mcclelland1987human}. 
\item[Need$ ^\text{\xmark}$] Represents what motivates and drives humans. For \citet{Maslow1943} needs are more than what is necessary for human survival. The chase to satisfy needs makes "man is a perpetually wanting animal." \citet{Maslow1943} represents five needs arranged in a hierarchy of prepotency. These are physiological, safety, love, esteem, and self-actualization. 
\end{description}
\label{box:teleo}
\end{concepts}

The \texttt{Value} class represents the teleological concept agents' intentions are aimed at. We chose (human) values to model meaning because this concept is (1) teleological, (2) operationalized, (3) shared  and (4) trans-situational. 

First, values can also be defined as `ideals worth pursuing' \citep{Dechesne2012}. Values are ends that people what to achieve: they are unreachable, but can be pursued \citep{Weide2011}. Values thus give reasons for agents to choose an action that contrasts with their habitual tendency. For example, by attaching the meaning efficiency to car-driving we can contrast the decision of an agent to take the car out of habit with the decision to intentionally take the train because it promotes environmentalism. Values thus provide the abstract aim intentions are directed at.

Second, \citet{Schwartz2012} developed several instruments (e.g. surveys) to operationalize values. Values thus provide an operational concept that has been made measurable and understandable in terms of empirical observations. This contrasts with the concept of `meaning' in SPT, where different authors wrote about different aspects of meaning. \citet{Schatzki1996} emphasized the 'teleoaffective' mental elements, such as embracing ends, purposes, projects and emotions. \citet{Reckwitz2002} focused on mental elements that are emotional or motivational. \citet{Shove2012} uses `meaning' as a bucket term that in addition comprises norms, ideas and aspirations. 
 
Third, the categorization of values by \citep{Schwartz2012} has been extensively empirically tested and shown to be consistent across 82 nations representing various age, cultural and religious groups \citep{Schwartz2012,Schwartz2012a,Bilsky2011,Davidov2008,Fontaine2008}. Values thus provide a shared concept that has a similar meanings to people around the world. For example, by relating car-driving with efficiency one gives it an aim that is understood in cultures all around the world. This satisfies our requirement to model a comprehensive set of social concepts.

Fourth, values are trans-situational \citep{Weide2011}. Values abstract over multiple situations and actions. This allows us to interconnect actions from different domains. For example, environmentalism can relate both to the practice of commuting as well as dining. This serves agent-based modellers that model individuals as part of a larger social systems over longer periods over time and, as such, comprise multiple actions that need to be related. Values contrasts with goals that focus on more concrete states within a domain and as such serve more precise co-ordination between agents. For example, car-commuting and hiking promote the same value (e.g., environmentalism), but not to the same goal (e.g., arriving at work, being active).

To enable heterogeneity between agents and the actions agents prefer we conceptualize the \texttt{ValuePriority} class and \texttt{ValueConnection} class. The \texttt{ValuePriority} class with an attribute \texttt{strength} represents how important an agent finds a value. For example, one agent finds the value of environmentalism more important than another agent. The \texttt{ValueConnection} class with the attribute \texttt{strength} represents how strongly an agent relates a value to an action. For example, one agent relates the value of fun to skateboarding while the other does not. The agent bases its intentional decision on strength of the \texttt{ValueConnection} class and \texttt{ValuePriority} class.  For example, an agent might value environmentalism highly and relate the activity \texttt{ride bike to work} to environmentalism, therefore, it chooses \texttt{ride bike to work} over \texttt{drive car to work}.  The \texttt{ValuePriority}, \texttt{ValueConnection} and \texttt{Value} class support modellers in modelling agents that direct their intentions at an abstract aim.


 \newpage
\section{Conceptualising Sociality}


SoPrA provides researchers with the primary concepts and relations to model agents that can make social decisions, update social variables an that reason about sociality (S2). Central to our conceptualisation of sociality is the connection between individual and social concepts. For example, norms, the concept referring to what the collective usually does, is connected to the concept of habits, the concept referring to what the individual usually does. Individual and social concepts are two sides of the same coin, or more explicit, two views on the same concept: a personal view and a collective view (S6, S7). In terms of views, habits are the personal view on what an agent usually does and norms are the collective view on what an agent usually does. The remainder of this section explains how we conceptualize a personal and collective view, link individual and collective concepts and model an implicit and explicit view. The resulting conceptualisation supports ABMs where agents use social practices to make social decisions such as coordinating their commute with their colleagues or partners.

\subsection{A Personal and Collective View}
\label{sec:conceptualising:sociality:pc}
SoPrA provides a personal and collective view on a social practice (S.3). Box \ref{box:pc} provides definitions of the concepts of common belief and collective intention that are relevant to conceptualize the \texttt{personalView}, \texttt{myCollectiveView} attributes that fulfills requirement S.3.

\begin{concepts}
\begin{description}
\item[Common Belief$^\simeq$] Everyone believes $\phi$ if all agents individually belief $\phi$. There is a common belief in $\phi$ if everyone believes in $\phi$, and also everyone believes everyone believes in $\phi$, etc. \citep{meyer2004epistemic}
\item[Collective Intention$^\simeq$]  Collective intentionality characterizes the intentionality that occurs when two or more individuals undertake a task together \citep{Searle1990,bratman1992shared}
\label{box:pc}
\end{description}
\end{concepts}

SoPrA uses the \texttt{personalView} attribute to refer to an agent’s personal view on a concept and the \texttt{myCollectiveView} to refer to an agent’s collective view on a concept. The \texttt{personalView} attribute specifies how strongly an agent itself believes an activity and an element are connected. The \texttt{myCollectiveView} attribute specifies how strong an agent believes the collective views (i.e, they believe) an activity and an element as connected. For example, in the \texttt{HabitualConnection} class the \texttt{personalView} attribute expresses how strong an agent believes that itself habitually connects a context element with an activity (e.g., an agent believes itself has a habit to use the car to commute). Analogously, the \texttt{myCollectiveView} attribute expresses how strong an agent believes others habitually connect a context element with an activity (e.g., how strong the agent believes that others habitually use their car to commute). The \texttt{personalView} and \texttt{myCollectiveView} attributes also feature in the \texttt{ChildParentRelation} class,  \texttt{ValueConnection} class and \texttt{ValuePriority} class.  The type of the view attributes is the same as the type of the main attribute in the class. Thus, in the \texttt{HabitualTrigger}, \texttt{ValueConnection} and \texttt{ValuePriority} class the type of the view attributes is an integer analogue to the \texttt{strength} attribute. In the \texttt{ChildParentRelation} class the type of the attribute is a string anologue to the \texttt{relationType} attribute (see Section \ref{sec:conceptualising:interconnectivity}). In short, the personal view and the collective view are captured in the \texttt{personalView} attribute and  \texttt{myCollectiveView} that feature in the \texttt{HabitualConnection}, \texttt{ChildParentRelation}, \texttt{ValueConnection} and \texttt{ValuePriority} class.

SoPrA ensures that agents are each able to have a different collective view (S9). This is emphasized by the `my' in \texttt{myCollectiveView}. This differs from the framework introduced in \citet{DignumSP} and formalized in \citet{DignumFormalSP}. \citet{DignumFormalSP} conceptualize an SP as \emph{one} collective entity. For example, there is \emph{the} SP of commuting instead of \emph{a} (view on the) SP of commuting. The SP in \citet{DignumFormalSP}’s framework is one that -- if agents choose to participate in the SP -- everyone believes (and everyone believes that everyone believes (i.e., a common belief, see Box 4.1). As such, \citet{DignumFormalSP}’s conceptualize social practices as entailing a collective intention.\footnote{Or in \citet{DignumFormalSP}'s words: a collective context that agents use to coordinate their actions.} conceptualising the SP as a single collective entity fits problems that focus on short-term action coordination where it's paramount all agents agree on the context (e.g., human-agent interaction). This is indeed one of the applications \citep{DignumFormalSP} has in mind  \citep{Augello2018}). SoPrA enables a fine-grained view where each agent is able to have a different collective view. This enables to model clusters of collective views on SPs (e.g., one social group views biking as a normal form of commuting whereas another does not). This fits problems studied using ABS: ABS researchers simulate the long-term dynamics of different views on the social word and their influence on both individuals and the social system. When necessary, SoPrA is easily adapted to match \citet{DignumFormalSP}'s framework by fixating the collective view (i.e., making the variable \texttt{myCollectiveView} static) and as such allowing only one collective view on an SP.
In short, SoPrA enables a fine-grained difference between collective views suitable for ABM and is easily extendable to support \emph{one} collective view whenever necessary (e.g., when applied to human-agent interaction).

SoPrA enables agents to have a different personal and collective view on an SP (S8). For example, an agent believes it personally has a habit to commute by skateboard but the collective has a habit to commute by car. This independence between views is conceptualized in the independence of the \texttt{personalView} and the \texttt{myCollectiveView} attribute. This way of separating the agent's personal view from its collective view differs from \citet{DignumFormalSP}'s framework. In \citet{DignumFormalSP}, agents are defined by concepts that are different from the concepts used to define an SP (e.g., a separation between habits and norms, goals and purpose, personal values and collective values). This separation between agents and SPs enables different types of agents to participate in the SP as long as a relation is specified between their concepts and the concepts used in the SP. Dignum thus aims for a meta-framework that enables plugging in agents specified in different languages  (e.g., BDI-agents, norm-based agents, Procedural Rule-Based agents). SoPrA aims for parsimony with an accessible computational implementation in ABM and therefore matches the language of the agent directly with the language of the SP. This facilitates both the parsimony of the model and the ability of an agent to make inferences between its personal and collective beliefs (S5). For example, an agent uses its own belief that commuting is usually done by car -- its car habit -- to infer that other people usually use a car to commute -- a car norm -- and asks another agent to carpool. SoPrA thus connects individual and social concepts to promote both the parsimony of the model and to supports agents in using the collective social world in individual decision-making.

\subsection{A Comprehensive Set of Social Concepts}
\label{sec:conceptualising:sociality:col}
By connecting individual and collective concepts SoPrA includes a comprehensive set of collective concepts (S3). Box \ref{box:compr} provides the definitions of a comprehensive set of collective concepts (i.e., norms, values, landmarks and purpose) that are either subsumed by our UML or not relevant for our requirements.

\begin{concepts}
\begin{description}
\item[Norms$^\simeq$] Norms generally refer to what is standard, acceptable or permissible behaviour in a group or society \citep{Fishbein2010}. 
\item[Values$^\text{\cmark}$] Represent what one finds important in life \citep{Vanderpoel2011}. For example, privacy, wealth or fairness. \citet{Schwartz2012} shows values represent labels that are collective (i.e, consistent across 82 nations).
\item[Landmarks$^\simeq$] Landmarks are states that represent states in a protocol where agents have a common belief about having reached that state \citep{Kumar2002}.
\item[Purpose$ ^\text{\xmark}$]  The purpose of an action is the reason for which the action is performed. The purpose is the common denominator over a sequence of actions and actors. For example, in the context of a lecture several actions (e.g, speaking or raising hands) done by several actors (e.g., students, lecturer) are all done for a common purpose (i.e., the students learn about the topic) \citep{DignumFormalSP}.
\end{description}
\label{box:compr}
\end{concepts}

SoPrA captures norms, values and collective activity associations. Norms were already treated as an example throughout this section as the collective counterpart of habits.  The \texttt{myCollectiveView} attribute in \texttt{HabitualConnection} class captures what an agent believes is the collective view on what actions usually occur given a certain context. This provides the basic ingredients for norms (i.e., in the terminology of Ostrom \citep{Ostrom2007} it contains the Actor, Aim and Condition elements of a norm). Values were treated as necessary as the teleological counterpart to habits (see Section \ref{sec:conceptualising:habits:teleo}). Section \ref{sec:conceptualising:habits:teleo} emphasized the personal view on values but values also have a collective meaning. Recall that the universality of values as demonstrated by Schwartz is one reason we chose values over other teleological concepts. SoPrA captures this collective view on values in the \texttt{myCollectiveView} attribute.  Activity associations are necessary to conceptualize interconnectivity and agents and also entail a collective view. Section \ref{sec:conceptualising:interconnectivity} treats this in more detail but for now it suffices to understand that a collective view on interconnectivity enables agents to understand how others believe activities are connected (corresponding to Landmarks (see Box \ref{box:compr}). For example, one believes others view car commuting as a kind-of commuting and therefore asks a colleague to carpool. Although rarely made explicit such beliefs are paramount to interpreting activities of others \citep{Metzinger2003, Okeya2014}. In short, by including a collective view on the individual concepts necessary to model habituality and interconnectivity SoPrA captures norms, values and collective activity associations. 

SoPrA provides a structure that enables a correct inclusion of additional collective concepts. For example, SoPrA provides the structure to correctly define more detailed statements about habits and norms. Currently, the \texttt{HabitualConnection} class captures a basic descriptive belief about what actions usually occur given a certain context. The \texttt{HabitualConnection} class is extendable by adding a \texttt{sanction} attribute. This enables a modeller to specify rules of the kind ``An agent may not drive a car in the city centre or else he will be fined” (see the MAIA framework \citep{Ghorbani2013}). The structure of SoPrA ensures such a \texttt{sanction} attribute entails both a personal view (e.g., I feel guilty when I drive) and a collective view (e.g., the collective fines me when I drive). Another example would be adding the concept of goals (as featured in \citet{DignumFormalSP}'s framework) and its collective counterpart: purpose (see Box). Goals and purpose provide teleological concepts that capture concrete reachable states of the world (as compared to the more abstract values) and are useful for applications of short-term human-agent interaction (as opposed to long-term ABM). Thus, SoPrA ensures that when a collective concept is added an individual counterpart is defined (and vice versa) such as adding `purpose’ requires a `goal’ and a ‘sanctioned norm’ requires a `sanctioned habit’.

\subsection{The Implicit and Explicit View}
SoPrA enables modellers to make a difference between the information implicit for the agents versus information explicit to the agent. For example, agents are able to have a strong implicit habit to interrupt people in conversation without having explicit awareness of this habit. As such, SoPrA enables differentiating agents that are able to only habitually decide to interrupt people from agents that are aware of this behaviour and able to change it. SoPrA captures this difference in a \texttt{strength} attribute -- an implicit strength -- and a \texttt{personalView} attribute -- an explicit habit strength. A difference in agents personal implicit strength and personal explicit view fits ABMs focus on behavioural change (e.g, tipping points only occur when there is an increased awareness of current habits). Besides, the provided structure -- that is, a difference between implicit and explicit variables -- is extendable to other variables. For example, an implicit collective view and explicit collective view fit applications in human-agent interaction. By making a difference between implicit and explicit information our model provides a structure to differentiate making decisions (based on implicit information) from reasoning (based on explicit information). 

This section showed how agents use a different personal and collective view to come to a set of social concepts that is used for social decision-making, updates and reasoning. This also clarifies the scope for social decision-making, updates and reasoning with social practices. For example, social practices are a heuristic used to make a difference between a personal and collective view and do not focus on chains of beliefs (e.g., I believe John believes I believe…) as represented in the `theory of mind’ theory\citep{Woodruff1978}. The next section treats the third aspect of SPs: interconnectivity.



\section{Conceptualising Interconnectivity}
\label{sec:conceptualising:interconnectivity} 
\begin{figure}[ht!]
\center
\includegraphics[width=\textwidth]{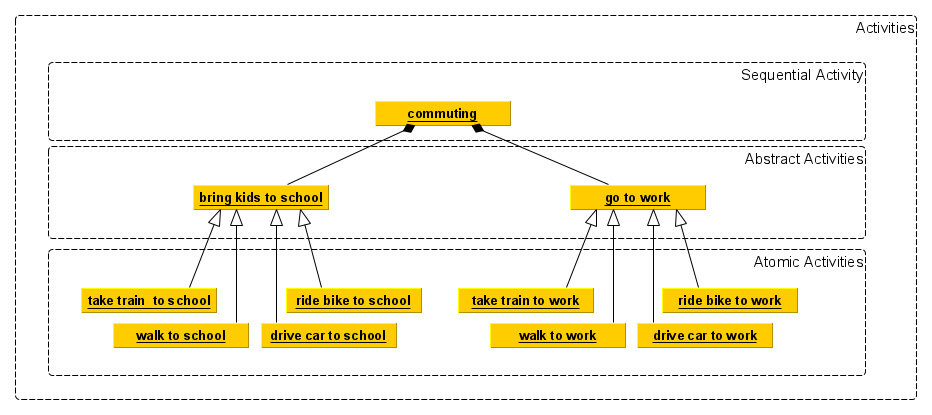}
\caption{An activity tree relating atomic activities, sequential activities and abstract activities for our use case of commuting. The white arrows specify an \texttt{ActivityConnection} of the type \texttt{isA} and the black diamonds an \texttt{ActivityConnection} of the type \texttt{partOf}}
\label{fig:activitytree}
\end{figure}

SoPrA provides researchers with the primary concepts and relations to model agents that can make interconnected decisions, update connections between activities and that reason about the interconnectedness of activities. Activities are connected \emph{indirectly} via time, space and common elements (I.3). These connections are captured in aspects of the model already treated. The \texttt{HabitualConnection} class connects multiple activities to the same location or time-points. The \texttt{ValueConnection} class connects multiple activities to the same element (i.e., values). In addition, SoPrA enables \emph{direct} connections between activities. These direct connections are featured in this section. They are temporal or ontological (I5) and connect different types of activities (I4). An example of different types of activities and their connections -- for the domain of community -- is found in Figure \ref{fig:activitytree}. The activity tree allows agents to, for example, reason that commuting has two parts (bringing the kids to school and going to work), reason that both parts are implementable using a car (i.e., drive car to school and drive car to work), and reason others believe commuting has two parts (and therefore borrow the car to your partner who needs it to bring the kids to school).  The remainder of this section will treat this activity hierarchy in more detail using Figure \ref{fig:activitytree} as an example. Box \ref{box:ic} provides the definitions of abstract actions, landmark states, Allen's logic and plan patterns that are relevant to conceptualize the \texttt{ActivityConnection} class that fulfills requirements I.1-I.5.

\begin{concepts}
\begin{description}
\item[Landmarks$^\simeq$] Landmarks represent states in a protocol where agents have a common belief about having reached that state \citep{Kumar2002}.
\item[Allen's Logic$^\simeq$] Allen's logic supports temporal relations that are relative, imprecise and uncertain \citep{Allen1983}. Allen's logic does not attribute a specific time $t$ to an event but the relation between two intervals. Give the intervals $t$ and $s$ some of examples are: $t$ before $s$, $t$ during $s$, $t$ equal to $s$ or $t$ overlaps with $s$.
\item[Plan Patterns$^\simeq$] Plan patterns describe usual patterns of actions defined by the landmarks that are expected to occur \citep{DignumFormalSP}.
\end{description}
\label{box:ic}
\end{concepts}

SoPrA uses the \texttt{activityType} attribute in the \texttt{Activity} class to make a difference between atomic activities, sequential activities and abstract activities (I4). Atomic activities represent concrete bodily movements that agents perform. In Figure \ref{fig:activitytree}, they are at the leaves of the activity tree, for example, \texttt{take train to school} is an atomic activity. Sequential activities are sequences of actions. In Figure \ref{fig:activitytree}, \texttt{commuting} is a sequential activity that sequences\texttt{bring kids to school} and \texttt{go to work}. Abstract activities are placeholders for different implementations of that activity. For example, \texttt{bring kids to school} is an abstract activity that is implementable by \texttt{take train to school}, \texttt{ride bike to school} or \texttt{walk to school}. Although abstract actions are not part of our requirement they are a consequence of the subsumption relation described in the next paragraph. These different types of activities enable modellers to model activities on different levels of abstractness and form the first part of the activity hierarchy.

The second part of the hierarchy consists of different connections between activities defined by the \texttt{ActivityConnection} class. This class has the attribute \texttt{relationType} that specifies two different types of relation between activities: the \texttt{isA} and \texttt{partOf} relation. The \texttt{isA} relation models the ontological subsumption relation between two activities. For example, taking the train to work is a kind of going to work. This is depicted as a white arrowhead in Figure \ref{fig:activitytree}. The \texttt{partOf} relation models a temporal relation between two activities: it specifies that all the child activities need to be completed to complete the parent activity. For example, car-commuting consists of bringing the kids to school and going to work.  Additional constructs for relating activities are found in \citet{DignumFormalSP} and Allen's Logic \citep{Allen1983}. Some are not required for our model because they suit coordination on a short time-scale (e.g., the parallel performance operator). Others are covered in our model by a combination of constructs. For example, a temporal sequence relation (i.e,. one action precludes the other) emerges (over time) in our model as a habitual connection between two activities instead of being fixed. SoPrA thus uses a \texttt{isA} and \texttt{partOf} relation to define the temporal and ontological connection necessary for ABM.

These different types of activities and different type of activity connection form an \emph{explicit} activity hierarchies. As mentioned in an earlier paper, this contrasts with BDI planning models where actities are a means to reach a certain state of the world (a goal) and the connections are \emph{implicit} in plans. By emphasizing the explicit relations between activities, we enable agents to reason about how these activities are connected. For example, by explicitly making the connection between bringing kids to school and driving a car to school an agent can reason about how bringing kids to school requires the resource car. Besides, the agent can ask their partner to bring their kids to school as they will have the same view on the connection between car commuting and bringing kids to school. Thus, by modelling explicit activity hierarchies we support both resource management and social reasoning about connections between activities.

\citet{DignumFormalSP}'s framework falls in between BDI agent and SoPrA in that it features activities in two ways: to reach a state as well as to connect activities. Explicit relations between activities feature in \citet{DignumFormalSP}'s framework in plan patterns. These plan patterns sequence abstract action much like in SoPrA, but in contrast to SoPrA and in line with BDI, \citet{DignumFormalSP} always mentions abstract actions in relation with a goal $\phi$ that the abstract action aims to achieve. This allows agents to have freedom in how to achieve an abstract action (i.e., how to achieve the goal phi related to the abstract action), while still keeping a certain rigidity by fixing how abstract actions are sequenced (as featured in the plan pattern). In comparison to SoPrA, this enables different types of agents to be plugged in. This is suitable for \citet{DignumFormalSP}’s purpose: the coordination of different agents in human-agent interaction but increases the complexity of their framework. SoPrA is aimed at ABM and therefore supports resource management and social reasoning while simplifying other parts to promote parsimony and clarity. 

Having an explicit and complete activity hierarchy provides agents with a structure they can use in making interconnected decisions.  For example, the decision-making of an agent starts at the activity at the top of the activity tree and goes step by step down the activity tree until it reaches an atomic activity. At each step, the agent uses the activity associations related to the candidate activities to make a decision. For example, it chooses go to work by car because it promotes efficiency. Or, it habitually chooses \texttt{walk to school} because there is a strong habitual trigger between the abstract activity \texttt{bring kids to school} and \texttt{walk to school} (recall that activities are also context-elements and thus habitually trigger other activities). After completing an action an agent returns either to the top of the activity tree (to make a decision at the next step) or to a sequential activity to complete both parts of the sequence. Or in a model featuring multiple social practices, the agent switches to another activity tree to sequence drinking a cup of coffee while keeping in mind the necessary actions related to commuting. The activity tree thus has a strong influence on the decisions agents make as it prescribes the possible actions and the possible decision paths in taking those actions. Therefore an ABS researcher should model the activity tree not only in accordance with empirical data but also the purpose of the study. For example, our example tree places great emphasis on bringing kids to school and going to work being two parts of commuting. This is useful to study the connection between the school-commute and work-commute but less suitable for other purposes.

\section{Discussion and Conclusion}
\label{sec:conceptualising:discussion}
This paper provides the domain-independent SoPrA framework that satisfies the requirements set out by \citep{Mercuur2020}. SoPrA provides an integration of habituality, sociality and interconnectivity in one agent model. SoPrA fills the three gaps in current agent models as identified by \citep{Mercuur2020}. First, it provides a framework that supports context-dependent habits, individual learning concerning habits and explicit reasoning about habits. Second, it provides a comprehensive set of collective concepts, orders social information around actions and relates individual and collective concepts in order to support agent interactions. Third, it defines relations between activities, between each activity and each other model concept and enables hierarchies of activities. All of the above allows modellers to simulate the dynamics of our routines. For example, \citet{Mercuur2015} used a preliminary version of SoPrA to show that an individual meat-eating habit more easily breaks in a new context (e.g., a vegetarian barbecue) and that such a habit, in turn, influences the collective view in the old context (i.e., your family having a more moderate view on vegetarians). The simulation shows that this results in a cascade of habit-breaking and, as such, a relatively effective intervention. By integrating habits, sociality and interconnectivity SoPrA extends current agent frameworks and as such enables new insights on our routines in complex systems.

This work builds upon the work by \citep{DignumFormalSP} but focuses on an agent framework for ABM instead of an agent framework for human-agent interaction. This is reflected in the choices presented throughout the paper. Most importantly, the choice to enable a connection between each activity, element and agent (Section \ref{sec:conceptualising:highlevelmodel}), the choice to use habitual triggers as a primary concept in the model (Section \ref{sec:conceptualising:habituality}), the choice to enable multiple collective views on an SP (Section \ref{sec:conceptualising:sociality:pc}) and the choice to use the same concepts to model the agent’s personal view and collective view (Section \ref{sec:conceptualising:sociality:col}). These choices result in a framework that supports ABM in simulating the long-term dynamics of complex systems. It enables modelling agents with habits and local collective views (e.g., one social group views biking as a normal form of commuting whereas another does not). Besides, by grouping certain concepts and using the same concepts to capture agents and SPs the framework is relatively accessible and parsimonious. For example, the concept of `actors’, `resources’, `places’, `start condition’, `duration’ are all captured in the \texttt{ContextElement} class and its relations to other concepts. Accessibility is, in particular, relevant for ABM, where social scientists aim to use computational tools. Parsimony is relevant for agent-based simulations (ABS) as this enables modellers to run more agents (representing larger groups of people) and keep the parameter space manageable. For applications where a more constraining framework is necessary SoPrA is extendable by fixing certain attributes. For example, by making the variable \texttt{myCollectiveView} static and as such allowing only one collective view on an SP. Thus, the choices made in this paper result in a parsimonious, accessible and flexible extension of \citep{DignumFormalSP} suited for ABM. 

SoPrA focusses on modularity to enhance its (re)useability for ABM. Modularization is a central desideratum in software engineering \citep{meyer1988object,ghezzi1991fundamentals,bergstra1990module,Riemsdijk2006}. By composing a system in modules, relatively independent units of functionality, we enhance its understandability and reuse. The modularity of SoPrA is reflected in its separation in classes and relations. This allows modellers to remove or reuse classes to fit the framework to his or her purpose. For example, removing the \texttt{ActivityConnection} class and restricting the model to atomic activities results in a model that emphasizes habituality and sociality (but not interconnectivity). Another useful purpose of modularity in ABM is to be able to trace back the results of a simulation to a particular module by switching part of the model on and off. This enables the modeller to check the validity of particular modules and gain clearer explanations via the simulation.

To further enhance the (re)useability, we aimed for compact modules, while at the same time, maintaining a correct, complete and consistent reflection of the requirements. Section \ref{sec:conceptualising:habituality}-\ref{sec:conceptualising:interconnectivity} have described the complete set of \citet{Mercuur2020}’s requirements and shown how they are correctly implemented in SoPrA. Appendix \ref{app:impl} describes our OWL framework that proves the consistency of SoPrA. We achieve the compactness of SoPrA in a process akin abstract painting where one depicts an object in a few on-point strokes of the brush. In modelling, one reorganizes and relabels concepts until the concepts form independent dimensions that correctly represent aspects of human behaviour. SoPrA uses three on-point dimensions: habituality, sociality and interconnectivity. Other complex concepts are reduceable to these dimensions. For example, sequenced behaviour is a combination of habituality and interconnectivity; a series of activities that trigger the next activity via a habitual trigger relation. A norm is a social version of a habitual trigger. A landmark action or plan pattern is a social version of an activation connection. In short, by focussing on three orthogonal dimensions and making grounded choices SoPrA is both correct, complete, consistent and compact.

Future work should focus on static or dynamic extensions of SoPrA. SoPrA is fit for our purpose -- modelling routines -- and meets our requirements. By connecting SoPrA with other modelling concepts (e.g., affordances, competences, roles, culture) one enables insights in the interaction of routines with other aspects of social systems. We foresee no problem in such an integration due to the modular and orthogonal nature of SoPrA. Appendix \ref{app:ext} provides a UML model that integrates SoPrA with additional agent concepts found in the literature and serves as a template for other desired extensions. Another avenue is to model the dynamic aspects of SPT by create a systemic translation from SPT to a domain-independent agent framework. This paper provides the static groundwork for such an extension.

The contribution of this paper is the SoPrA framework that provides ABM researchers with a means to simulate our routines. The language we use in SoPrA illuminates the habitual, social and interconnected aspects of social systems via simulation. This means we enable a new way to know and explore the world and gain new insights. These insights lead to new ways to control and improve our world. The view we enable is one of empirical and theoretical importance. Our routines have been emphasized since the old greeks and can now be combined with the strengths of computer simulation. We envision that this allows insights into the interplay of habits and norms; for instance, can we understand the slow change in climate behaviour as a consequence of the interplay of group inertia (due to norms) and individual inertia (due to habits)?  How to best influence multiple societal groups using the dynamics of individual habits and local norms to our advantage? Are these behaviours intrinsically connected with other behaviours through routines and do they need to be targeted as such; what does changing one of these behaviours mean for the others? SoPrA provides the language to formulate these questions in precise terms. Furthermore, this better formulation is supported by SoPrA that allows simulation dedicated to exploring these questions.


\newpage
\appendix
\section{Overview Requirements}
\label{app:req}
\renewcommand{\arraystretch}{1.7}
\rowcolors{2}{gray!25}{white}

Table \ref{tab:my-table} provides an overview on the requirements on habituality sociality and interconnectivity.

\begin{table}[ht!]
\centering
\caption{An overview on the requirements on habituality sociality and interconnectivity}
\label{tab:my-table}
\begin{tabular}{@{}C{0.6cm}C{5cm}C{5cm}C{5cm}@{}}
\toprule
\rowcolor[HTML]{FFFFFF} 
 & \multicolumn{3}{c}{\cellcolor[HTML]{FFFFFF}We require the model to...} \\
 \cmidrule(r{2pt}){1-1} \cmidrule(l){2-2} \cmidrule(l){3-3} \cmidrule(l){4-4}
\rowcolor[HTML]{FFFFFF} 
Nr. & Habituality & Sociality & Interconnectivity \\ \midrule
1 & capture that social practices are habitual as they are similar over time & capture that social practices are social as they are similar over people. & capture that social practices are interconnected as they are similar over activities. \\
2 & provide the primary concepts and relations to researchers to model habitual decisions, updates and reasoning & provide the primary concepts and relations to enable agents to make socially intelligent decisions, update social information and reason about collective concepts. & provide the primary concepts and relations to enable agents to make interconnected decisions, updates and reason about the interconnectedness of activities. \\
3 & enable agents to differentiate between habits and intentions & use a comprehensive set of collective concepts that support social decision-making, updating and reasoning. & express that social practices are connected in terms of time, space and common elements. \\
4 & support habitual relations between an action and a context-element where the strength of that relationship is a continuous parameter. & support agents that order social information around their practices. & differentiate between different types of activities: atomic activities and sequential activities (an ordered sequence of actions). \\
5 & capture that context-elements can comprise resources, activities, location, timepoints or other people. & capture that social practices relate the collective social world with the individual world of interactions. & capture both temporal and ontological relations between activities to enable agents to make decisions and inferences. \\
6 & to capture that agents can differ in the strength of a connection, maximum strength, time to reach this maximum, amount of attention they attribute to an action & capture that agents have a personal view on a social practice. &  \\
7 & to provide a concept that captures the abstract aim intentions are directed at. & capture that agents have a collective view on a social practice. &  \\
 &  & enable agents to have a different personal view than their collective view. &  \\
 &  & enable agents to each have a different collective view on a social practice. &  \\ \bottomrule
\end{tabular}
\end{table}

\section{Extensions}
\label{app:ext}
\begin{figure}[ht]
    \centering
    \includegraphics[width=\textwidth]{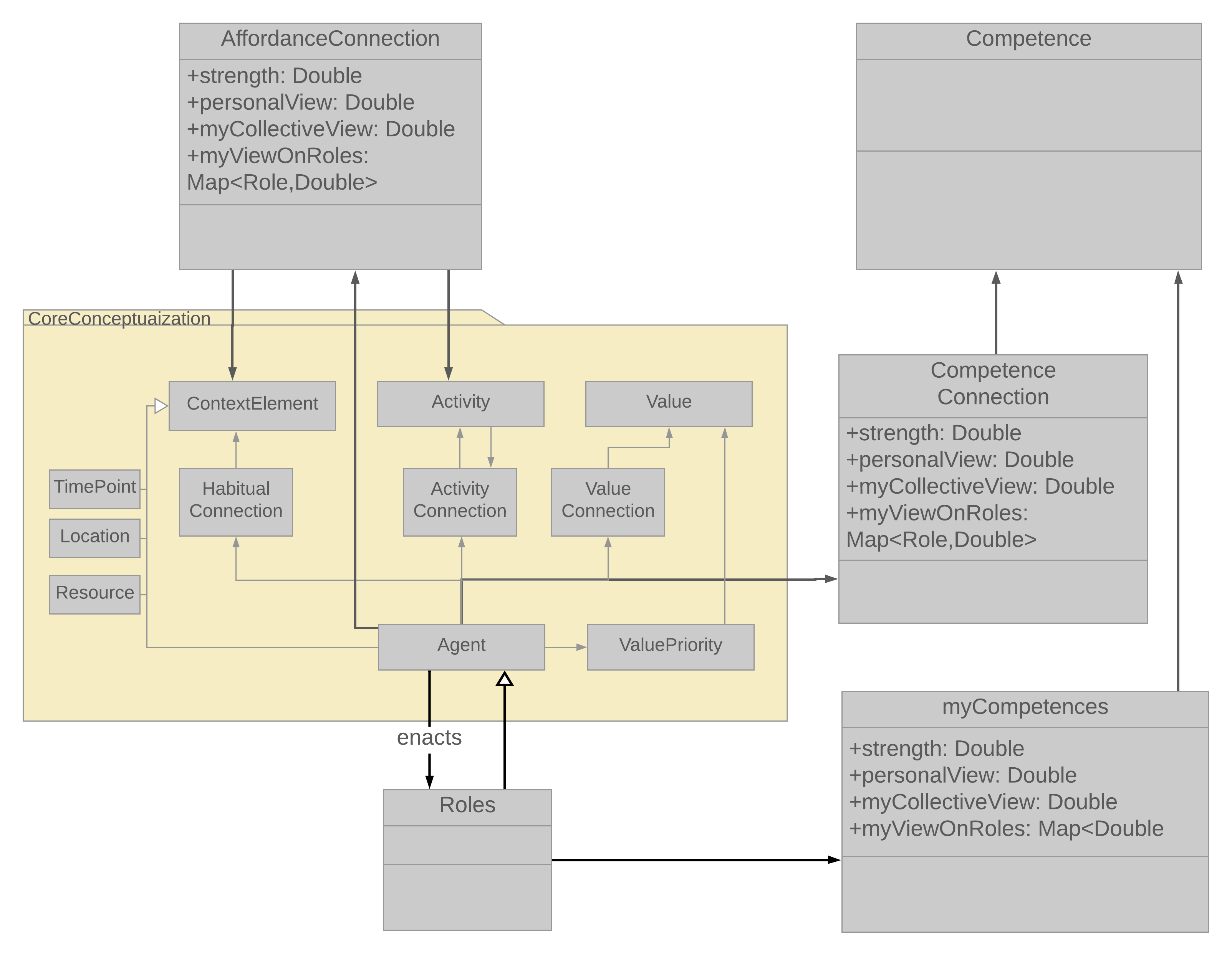}
    \caption{A UML diagram that extends the core SoPrA framework adding affordances, competences and roles.}
    \label{fig:ext}
\end{figure}

Figure \ref{fig:ext} provides an extension of our core SoPrA framework. We treat concepts that are mentioned in related work but are not necessary to fulfil our requirements. 

\begin{description}
\item[Roles] Roles enable modellers to group similar agents. Grouping agents has several advantages. First, instead of defining multiple agents that have the same beliefs (e.g., car driving is efficient), modellers can define a role (e.g., a parent) that believes car driving is efficient. Figure \ref{fig:ext} reflects the possibility for the \texttt{Role} class to have the same connections as the \texttt{Agent} class with the generalization arrow. Connecting a belief once to a role instead of connecting multiple beliefs to multiple agents reduces the spatial complexity of the model. Second, roles enable agents to enact multiple roles and thus entertain multiple views. This is reflected in the \texttt{enact} association from the \texttt{Agent} class to the \texttt{Role} class. Third, roles enable an agent to hold a view on a specific group of agents. This feature is reflected in the attribute \texttt{myViewOnRoles} in the classes that connect agents and elements. This attribute maps a \texttt{Role} set (i.e., groups of agents) to a \texttt{Double} representing how strongly the agent believes this group of agents agrees on the connection (e.g., parents believe car driving is efficient, parents believe car-driving requires the competence to drive a car).\footnote{Although this attribute enables the modeller to specify beliefs about groups of agents for \emph{any} connection (e.g., a habitual connection, a value connection), for ease of presentation Figure \ref{fig:ext} only depicts this attribute in the classes related to the extensions .}
\item[Affordances] Affordances specify for a context-element what actions it makes possible \citep{gibson1979ecological}. For example, a chair affords to sit. Affordances are related to habitual triggers. Affordances specify what actions are possible, habitual triggers which actions are salient \citep{DignumFormalSP}. Our extended framework contains the \texttt{AffordanceConnection} class as an activity association analogue to habitual triggers. Specifying the \texttt{strength} of the connection as a \texttt{Double} enables both a more restricted view on affordances (e.g., a chair affords sitting so the strength is 1, but a table does not so the strength is 0) as well as a continuous view on affordances (e.g., a chair is better for sitting than a table). 
\item[Competences] A competence is the capability to make a certain proposition true~\citep{Dignum2012}. Competencies are criteria within the agent that enable an action (whereas affordances are criteria outside the agent that enable an action). In \citet{Shove2012} competencies represent the implicit skill or know-how that is needed to act. In SoPrA-extended competences are added as both an activity association (i.e., \texttt{CompetenceConnection}) and an agent association (i.e., \texttt{myCompetences}) analogue to values. Specifying the \texttt{strength} of the connection as a double enables both a more restricted view on competences (e.g., a driver needs to know how to stay calm to drive) as well as a continuous view on competences (e.g., it is better when a driver knows how to stay calm when driving).
\end{description}

\section{Formal and Computational versions of SoPrA}
\label{app:impl}
SoPrA has been implemented in Repast Simphony, and Prot\'eg\'e. The code is available at Github.\footnote{An implementation of SoPrA is available at https://github.com/rmercuur/SoPrA-Models}

\textbf{Repast Simphony} Repast Simphony is a java-based tool for programming ABS based on the Eclipse IDE \citep{North2013}. The translation from the UML to java code is relatively straight-forward as both are based on the object-orientated (OO) design method. Our implementation uses the \texttt{Table} class from Google's Guava to simplify the classes in the \texttt{activityAssociation} and \texttt{agentAssociation} package. For example, the \texttt{HabitualConnection} class is transformed into a table in the agent class that maps a \texttt{ContextElement} and \texttt{Activity} pair to three doubles: the habitual strength, the personal view on the habitual strength and the collective view on the habitual strength (e.g.,  \texttt{Table<ContextElement, Activity, StrengthValues<Double, Double, Double>}). Our implementation comes with an example simulation that uses SoPrA to simulate the dynamics of commuting behaviour \citep{MercuurDiscerning2020}. Because both UML and Repast are based on OO-desisgn Repast is the recommend ABM system to use SoPrA.

\textbf{OWL Prot\'eg\'e} Our online repository provides a formal framework in the OWL-language. A formal framework promotes unambiguity, rigour and enables formal reasoning. Prot\'eg\'e is a tool to author formal ontologies and enable formal reasoners that aid the modeller \citep{Protege}.  Using Protege –a formal reasoner – we show the framework is consistent, verify our domainmodels correctly implement the framework and verify properties.
In detail, Prot\'eg\'e enables modellers to make domain-specific model and use the formal reasoner to check if these models corresponds to the domain-independent framework we made. Prot\'eg\'e uses description logic: a logic that trades its expressivity to acquire decidability \citep{Horrocks2007}. The UML classes correspond to classes in Prot\'eg\'e and UML associations to object properties. \citet{Mercuur2018DL} describes the translation process from (an earlier version of SoPrA) to description logic. This shows that description logic is expressive enough to capture the semantics of SoPrA and that the formalization of SoPrA in Prot\'eg\'e is satisfiable. Prot\'eg\'e can be used in combination with ABM to infer new knowledge for the agents. For example, one can use the formal reasoner to apply to automatically infer that if riding a bike, taking the train to work and walking promote environmentalism then non-car commuting also promotes environmentalism. In short, researchers who want to verify that their domain model corresponds to SoPrA or want to infer new knowledge are recommended to use Prot\'eg\'e.

\bibliographystyle{splncsnat}
\bibliography{Mendeley.bib}

\begin{thebibliography}{58}
\providecommand{\natexlab}[1]{#1}
\providecommand{\url}[1]{\texttt{#1}}
\providecommand{\urlprefix}{}

\bibitem[{Allen(1983)}]{Allen1983}
Allen, J.: {Maintaining knowledge about temporal intervals}.
\newblock Communications of the ACM 26(11), 832--843 (1983)

\bibitem[{Anderson(2005)}]{Anderson2005}
Anderson, J.R.: {Cognitive Psychology an its implications}.
\newblock Macmillan (2005)

\bibitem[{Augello et~al.(2018)Augello, Dignum, Gentile, Infantino, Maniscalco,
  Pilato, and Vella}]{Augello2018}
Augello, A., Dignum, F., Gentile, M., Infantino, I., Maniscalco, U., Pilato,
  G., Vella, F.: {A social practice oriented signs detection for human-humanoid
  interaction}.
\newblock Biologically Inspired Cognitive Architectures 25(June), 8--16 (2018),
  \urlprefix\url{https://doi.org/10.1016/j.bica.2018.07.013}

\bibitem[{Bergstra et~al.(1990)Bergstra, Heering, and
  Klint}]{bergstra1990module}
Bergstra, J.A., Heering, J., Klint, P.: {Module algebra}.
\newblock Journal of the ACM (JACM) 37(2), 335--372 (1990)

\bibitem[{Bersini(2014)}]{Bersini2012}
Bersini, H.: {UML for ABM}.
\newblock Journal of Artificial Societies and Social Simulation 15(2012), 1--16
  (2014)

\bibitem[{Bilsky et~al.(2011)Bilsky, Janik, and Schwartz}]{Bilsky2011}
Bilsky, W., Janik, M., Schwartz, S.H.: {The Structural Organization of Human
  Values-Evidence from Three Rounds of the European Social Survey (ESS)}.
\newblock Journal of Cross-Cultural Psychology 42(5), 759--776 (2011)

\bibitem[{Bourdieu(1977)}]{Bourdieau1977}
Bourdieu, P.: {Outline of a theory of practice}, vol.~16.
\newblock Cambridge University Press (1977),
  \urlprefix\url{papers://cfc50b6a-2d9e-4feb-87e5-d6012043bd5a/Paper/p2118}

\bibitem[{Bratman(1992)}]{bratman1992shared}
Bratman, M.E.: {Shared cooperative activity}.
\newblock The philosophical review 101(2), 327--341 (1992)

\bibitem[{Crawford and Ostrom(2007)}]{Ostrom2007}
Crawford, S.E.S., Ostrom, E.: {A Grammar of Institutions}.
\newblock Political Science 89(3), 582--600 (2007),
  \urlprefix\url{http://www.jstor.org/stable/2082975}

\bibitem[{Davidov et~al.(2008)Davidov, Schmidt, and Schwartz}]{Davidov2008}
Davidov, E., Schmidt, P., Schwartz, S.H.: {Bringing values back in: The
  adequacy of the European social survey to measure values in 20 countries}.
\newblock Public Opinion Quarterly 72(3), 420--445 (2008)

\bibitem[{Dechesne et~al.(2012)Dechesne, Di~Tosto, Dignum, and
  Dignum}]{Dechesne2012}
Dechesne, F., Di~Tosto, G., Dignum, V., Dignum, F.: {No Smoking Here: Values,
  Norms and Culture in Multi-Agent Systems}.
\newblock Artificial Intelligence and Law 21(1), 79--107 (8 2012),
  \urlprefix\url{http://link.springer.com/10.1007/s10506-012-9128-5}

\bibitem[{Dey(2001)}]{Dey2001}
Dey, A.K.: {Understanding and using context}.
\newblock Personal and Ubiquitous Computing 5(1), 4--7 (2001)

\bibitem[{Dignum(2018)}]{DignumFormalSP}
Dignum, F.: {Interactions as Social Practices: towards a formalization}.
\newblock ArXiv e-prints pp. 1--36 (2018),
  \urlprefix\url{http://arxiv.org/abs/1809.08751}

\bibitem[{Dignum and Dignum(2012)}]{Dignum2012}
Dignum, V., Dignum, F.: {A logic of agent organizations}.
\newblock Logic Journal of IGPL 20(September), 3--7 (2012),
  \urlprefix\url{http://jigpal.oxfordjournals.org/content/20/1/283.short}

\bibitem[{Dignum and Dignum(2015)}]{DignumSP}
Dignum, V., Dignum, F.: {Contextualized Planning Using Social Practices}.
\newblock In: Ghose, A., Oren, N., Telang, P., Thangarajah, J. (eds.)
  Coordination, Organizations, Institutions, and Norms in Agent Systems, pp.
  36--52. Springer International Publishing, Cham (2015)

\bibitem[{Fishbein and Azjen(2011)}]{Fishbein2010}
Fishbein, M., Azjen, I.: {Predicting and Changing Behavior: The Reasoned Action
  Approach}.
\newblock Taylor {\&} Francis (2011)

\bibitem[{Fontaine et~al.(2008)Fontaine, Poortinga, Delbeke, and
  Schwartz}]{Fontaine2008}
Fontaine, J.R.J., Poortinga, Y.H., Delbeke, L., Schwartz, S.H.: {Structural
  Equivalence of the Values Domain Across Cultures: Distinguishing Sampling
  Fluctuations From Meaningful Variation}.
\newblock Journal of Cross-Cultural Psychology 39(October), 345--365 (2008)

\bibitem[{Ghezzi et~al.(1991)Ghezzi, Jazayeri, and
  Mandrioli}]{ghezzi1991fundamentals}
Ghezzi, C., Jazayeri, M., Mandrioli, D.: {Fundamentals of software
  engineering}.
\newblock Prentice Hall Englewood Cliffs (1991)

\bibitem[{Ghorbani et~al.(2013)Ghorbani, Bots, Dignum, and
  Dijkema}]{Ghorbani2013}
Ghorbani, A., Bots, P., Dignum, V., Dijkema, G.: {MAIA: A framework for
  developing agent-based social simulations}.
\newblock Jasss 16(2), 1--15 (2013)

\bibitem[{Gibson(1979)}]{gibson1979ecological}
Gibson, J.J.: {The ecological approach to visual perception. Boston, MA, US}
  (1979)

\bibitem[{Holtz(2014)}]{Holtz2014}
Holtz, G.: {Generating Social Practices}.
\newblock Journal of Artificial Societies and Social Simulation 17(1), 17
  (2014)

\bibitem[{Horrocks(2007)}]{Horrocks2007}
Horrocks, I.: {Description Logic : A Formal Foundation for Ontology Languages
  and Tools} (2007)

\bibitem[{Kumar et~al.(2002)Kumar, Huber, Cohen, and McGee}]{Kumar2002}
Kumar, S., Huber, M.J., Cohen, P.R., McGee, D.R.: {Toward a formalism for
  conversation protocols using joint intention theory}.
\newblock Computational Intelligence 18(2), 174--228 (2002)

\bibitem[{Latour(1996)}]{Latour1996}
Latour, B.: {On actor-network theory: A few clarifications} (1996),
  \urlprefix\url{https://www.jstor.org/stable/40878163}

\bibitem[{Locke and Latham(1990)}]{locke1990theory}
Locke, E.A., Latham, G.P.: {A theory of goal setting {\&} task performance.}
\newblock Prentice-Hall, Inc (1990)

\bibitem[{MASLOW(1943)}]{Maslow1943}
MASLOW, A.H.: {A theory of human motivation}.
\newblock Psychological Review  (1943)

\bibitem[{McClelland(1987)}]{mcclelland1987human}
McClelland, D.C.: {Human motivation}.
\newblock CUP Archive (1987)

\bibitem[{Mercuur(2015)}]{Mercuur2015}
Mercuur, R.: {Interventions on Contextualized Decision Making : an Agent-Based
  Simulation Study}.
\newblock Ph.D. thesis, Utrecht University (2015),
  \urlprefix\url{https://dspace.library.uu.nl/handle/1874/323482}

\bibitem[{Mercuur et~al.(2017)Mercuur, Dignum, and Jonker}]{Mercuur2017}
Mercuur, R., Dignum, V., Jonker, C.: {Modeling Social Preferences with Values}.
\newblock In: The Proceedings of the 3nd International Workshop on Smart
  Simulation and Modelling for Complex Systems @ IJCAI. pp. 17--28 (2017)

\bibitem[{Mercuur et~al.(2020)Mercuur, Dignum, and Jonker}]{Mercuur2020}
Mercuur, R., Dignum, V., Jonker, C.: {Integrating Social Practice Theory in
  Agent-Based Models: A Review of Theories and Agents}.
\newblock IEEE Transactions on Computational Social Systems  (2020)

\bibitem[{Mercuur et~al.(2019)Mercuur, Dignum, and
  Jonker}]{MercuurDiscerning2020}
Mercuur, R., Dignum, V., Jonker, C.M.: {Discerning Between Two Theories on
  Habits with Agent-based Simulation}.
\newblock In: Eggert, N. (ed.) Proceedings of the Social Simulation Conference
  2019. p. (in press). Springer LNCS, Mainz (2019)

\bibitem[{Mercuur et~al.(2018)Mercuur, Larsen, and Dignum}]{Mercuur2018DL}
Mercuur, R., Larsen, J.B., Dignum, V.: {Modelling the Social Practices of an
  Emergency Room to Ensure Staff and Patient Wellbeing}.
\newblock In: 30th Benelux Conference on Artificial Intelligence. pp. 133--147.
  Den Bosch (2018)

\bibitem[{Metzinger and Gallese(2003)}]{Metzinger2003}
Metzinger, T., Gallese, V.: {The emergence of a shared action ontology:
  Building blocks for a theory}.
\newblock Consciousness and Cognition 12(4), 549--571 (2003)

\bibitem[{Meyer(1988)}]{meyer1988object}
Meyer, B.: {Object-oriented software construction}, vol.~2.
\newblock Prentice hall New York (1988)

\bibitem[{Meyer and Van Der~Hoek(2004)}]{meyer2004epistemic}
Meyer, J.J.C., Van Der~Hoek, W.: {Epistemic logic for AI and computer science},
  vol.~41.
\newblock Cambridge University Press (2004)

\bibitem[{Musen(2015)}]{Protege}
Musen, M.A.: {The prot{\'{e}}g{\'{e}} project: a look back and a look forward}.
\newblock {\{}AI{\}} Matters 1(4), 4--12 (2015),
  \urlprefix\url{https://doi.org/10.1145/2757001.2757003}

\bibitem[{Narasimhan et~al.(2017)Narasimhan, Roberts, Xenitidou, and
  Gilbert}]{Narasimhan2017}
Narasimhan, K., Roberts, T., Xenitidou, M., Gilbert, N.: {Using ABM to Clarify
  and Refine Social Practice Theory}.
\newblock In: Jager, W. (ed.) Advances in Social Simulation 2015. vol. 528.
  Springer International Publishing AG 2017 (2017),
  \urlprefix\url{http://link.springer.com/10.1007/978-3-319-47253-9}

\bibitem[{Neal et~al.(2012)Neal, Wood, Labrecque, and Lally}]{Neal2012}
Neal, D.T., Wood, W., Labrecque, J.S., Lally, P.: {How do habits guide
  behavior? Perceived and actual triggers of habits in daily life}.
\newblock Journal of Experimental Social Psychology 48(2), 492--498 (2012),
  \urlprefix\url{http://dx.doi.org/10.1016/j.jesp.2011.10.011}

\bibitem[{North et~al.(2013)North, Collier, Ozik, Tatara, Macal, Bragen, and
  Sydelko}]{North2013}
North, M.J., Collier, N.T., Ozik, J., Tatara, E.R., Macal, C.M., Bragen, M.,
  Sydelko, P.: {Complex adaptive systems modeling with Repast Simphony}.
\newblock Complex Adaptive Systems Modeling 1(1), 3 (2013),
  \urlprefix\url{http://casmodeling.springeropen.com/articles/10.1186/2194-3206-1-3}

\bibitem[{Okeyo et~al.(2014)Okeyo, Chen, and Wang}]{Okeya2014}
Okeyo, G., Chen, L., Wang, H.: {Combining ontological and temporal formalisms
  for composite activity modelling and recognition in smart homes}.
\newblock Future Generation Computer Systems 39, 29--43 (2014)

\bibitem[{Poel and Royakkers(2011)}]{Vanderpoel2011}
Poel, I.v.d., Royakkers, L.: {Ethics, Technology, and Engineering: An
  Introduction}.
\newblock John Wiley {\&} Sons (2011),
  \urlprefix\url{http://eu.wiley.com/WileyCDA/WileyTitle/productCd-EHEP002302.html}

\bibitem[{Rao and Georgeff(1995)}]{BDI}
Rao, A., Georgeff, M.: {BDI Agents: From Theory to Practice}.
\newblock Proceedings of the First International Conference on Multi-Agent
  Systems (ICMAS-95)  (1995)

\bibitem[{Reckwitz(2002)}]{Reckwitz2002}
Reckwitz, A.: {Toward a theory of social practices: A development in
  culturalist theorizing}.
\newblock European Journal of Social Theory 5(2), 243--263 (2002)

\bibitem[{Schatzki(1996)}]{Schatzki1996}
Schatzki, T.R.: {Social Practices. A Wittgensteinian approach to human activity
  and the social}.
\newblock Cambridge University Press (1996)

\bibitem[{Schwartz(2012)}]{Schwartz2012}
Schwartz, S.H.: {An Overview of the Schwartz Theory of Basic Values}.
\newblock Online Readings in Psychology and Culture 2, 1--20 (2012)

\bibitem[{Schwartz et~al.(2012)Schwartz, Cieciuch, Vecchione, Davidov, Fischer,
  Beierlein, Ramos, Verkasalo, L{\"{o}}nnqvist, Demirutku, Dirilen-Gumus, and
  Konty}]{Schwartz2012a}
Schwartz, S.H., Cieciuch, J., Vecchione, M., Davidov, E., Fischer, R.,
  Beierlein, C., Ramos, A., Verkasalo, M., L{\"{o}}nnqvist, J.E., Demirutku,
  K., Dirilen-Gumus, O., Konty, M.: {Refining the theory of basic individual
  values.}
\newblock Journal of Personality and Social Psychology 103(4), 663--688 (2012)

\bibitem[{Searle(1990)}]{Searle1990}
Searle, J.: {Collective intentions and actions}.
\newblock Intentions in communication p. 401 (1990)

\bibitem[{Searle(1995)}]{Searle1995}
Searle, J.: {The Construction of Social Reality} (1995)

\bibitem[{Shove et~al.(2012)Shove, Pantzar, and Watson}]{Shove2012}
Shove, E., Pantzar, M., Watson, M.: {The Dynamics of Social Practice: Everyday
  Life and How it Changes}.
\newblock SAGE Publications, London (2012)

\bibitem[{Shove et~al.(2015)Shove, Watson, and Spurling}]{Shove2015}
Shove, E., Watson, M., Spurling, N.: {Conceptualizing connections: Energy
  demand, infrastructures and social practices}.
\newblock European Journal of Social Theory 18(3), 274--287 (2015),
  \urlprefix\url{http://journals.sagepub.com/doi/10.1177/1368431015579964}

\bibitem[{Taylor(1973)}]{Taylor1973}
Taylor, C.: {Interpretation and the Sciences of Man}.
\newblock In: Explorations in Phenomenology, vol.~25, pp. 47--101. Philosophy
  Education Society Inc. (1973)

\bibitem[{Van~Riemsdijk et~al.(2006)Van~Riemsdijk, Dastani, Meyer, and
  De~Boer}]{Riemsdijk2006}
Van~Riemsdijk, M.B., Dastani, M., Meyer, J.J.C., De~Boer, F.S.: {Goal-oriented
  modularity in agent programming}.
\newblock Proceedings of the International Conference on Autonomous Agents
  2006, 1271--1278 (2006)

\bibitem[{Verplanken et~al.(1994)Verplanken, Aarts, van Knippenberg, and van
  Knippenberg}]{Verplanken1994}
Verplanken, B., Aarts, H., van Knippenberg, A., van Knippenberg, C.: {Attitude
  Versus General Habit: Antecedents of Travel Mode Choice}.
\newblock Journal of Applied Social Psychology 24(4), 285--300 (1994)

\bibitem[{Verplanken and Orbell(2003)}]{Verplanken2003}
Verplanken, B., Orbell, S.: {Reflections on Past Behavior: A Self-Report Index
  of Habit Strength}.
\newblock Journal of Applied Social Psychology 33(6), 1313--1330 (2003)

\bibitem[{Weide(2011)}]{Weide2011}
Weide, T.v.d.: {Arguing to motivate decisions}.
\newblock SIKS Dissertation Series  (2011),
  \urlprefix\url{http://www.narcis.nl/publication/Language/EN/id/170/RecordID/oai:dspace.library.uu.nl:1874/210788}

\bibitem[{Wetzel(2018)}]{sep-types-tokens}
Wetzel, L.: {Types and Tokens}.
\newblock In: Zalta, E.N. (ed.) The Stanford Encyclopedia of Philosophy.
  Metaphysics Research Lab, Stanford University, fall 2018 edn. (2018)

\bibitem[{Wood et~al.(2002)Wood, Quinn, and Kashy}]{Wood2002}
Wood, W., Quinn, J.M., Kashy, D.A.: {Habits in everyday life: Thought, emotion,
  and action}.
\newblock Journal of Personality and Social Psychology 83(6), 1281--1297 (2002)

\bibitem[{Woodruff(1978)}]{Woodruff1978}
Woodruff, G.: {Premack and Woodruff : Chimpanzee theory of mind}.
\newblock Behavioral and Brain Sciences 1(1978), 515--526 (1978)

\end{thebibliography}
\end{document}